\documentclass[letterpaper,preprint]{aastex}

\usepackage{amsmath}
\usepackage{booktabs}
\usepackage{braket}
\usepackage{graphicx}
\usepackage{lineno}
\usepackage[margin=1in]{geometry}
\usepackage{units}

\usepackage{natbib}
\bibpunct{(}{)}{;}{a}{}{,}

\newcommand{\Gp}[1] {\ensuremath{\left(#1\right)}}
\newcommand{\Gb}[1] {\ensuremath{\left[#1\right]}}
\newcommand{\GB}[1] {\ensuremath{\left\{#1\right\}}}
\newcommand{\sci}[2]{\ensuremath{{#1}\times10^{#2}}}

\newcommand*\patchAmsMathEnvironmentForLineno[1]{%
  \expandafter\let\csname old#1\expandafter\endcsname\csname #1\endcsname
  \expandafter\let\csname oldend#1\expandafter\endcsname\csname end#1\endcsname
  \renewenvironment{#1}%
     {\linenomath\csname old#1\endcsname}%
     {\csname oldend#1\endcsname\endlinenomath}}%
\newcommand*\patchBothAmsMathEnvironmentsForLineno[1]{%
  \patchAmsMathEnvironmentForLineno{#1}%
  \patchAmsMathEnvironmentForLineno{#1*}}%
\AtBeginDocument{%
\patchBothAmsMathEnvironmentsForLineno{equation}%
\patchBothAmsMathEnvironmentsForLineno{align}%
\patchBothAmsMathEnvironmentsForLineno{flalign}%
\patchBothAmsMathEnvironmentsForLineno{alignat}%
\patchBothAmsMathEnvironmentsForLineno{gather}%
\patchBothAmsMathEnvironmentsForLineno{multline}%
}

\title{Search for Prompt Neutrino Emission from Gamma-Ray Bursts with
IceCube}

\shortauthors{M.~G.~Aartsen et al.}
\author{
IceCube Collaboration:
M.~G.~Aartsen\altaffilmark{1},
M.~Ackermann\altaffilmark{2},
J.~Adams\altaffilmark{3},
J.~A.~Aguilar\altaffilmark{4},
M.~Ahlers\altaffilmark{5},
M.~Ahrens\altaffilmark{6},
D.~Altmann\altaffilmark{7},
T.~Anderson\altaffilmark{8},
C.~Arguelles\altaffilmark{5}
% ...
}
\altaffiltext{1}{School of Chemistry \& Physics, University of Adelaide, Adelaide SA, 5005 Australia}
\altaffiltext{2}{DESY, D-15735 Zeuthen, Germany}
\altaffiltext{3}{Dept.~of Physics and Astronomy, University of Canterbury, Private Bag 4800, Christchurch, New Zealand}
\altaffiltext{4}{Universit\'e Libre de Bruxelles, Science Faculty CP230, B-1050 Brussels, Belgium}
\altaffiltext{5}{Dept.~of Physics and Wisconsin IceCube Particle Astrophysics Center, University of Wisconsin, Madison, WI 53706, USA}
\altaffiltext{6}{Oskar Klein Centre and Dept.~of Physics, Stockholm University, SE-10691 Stockholm, Sweden}
\altaffiltext{7}{Erlangen Centre for Astroparticle Physics, Friedrich-Alexander-Universit\"at Erlangen-N\"urnberg, D-91058 Erlangen, Germany}
\altaffiltext{8}{Dept.~of Physics, Pennsylvania State University, University Park, PA 16802, USA}
% ...

\date{}

\begin{abstract}
  We present constraints derived from a search of four years of IceCube data
  for a prompt neutrino flux from gamma-ray bursts (GRBs).  A single
  low-significance neutrino, compatible with the atmospheric neutrino
  background, was found in coincidence with one of the 506 observed bursts.
  Although GRBs have been proposed as candidate sources for ultra-high
  energy cosmic rays, our limits on the neutrino flux disfavor much of the
  parameter space for the latest models.  We also find that no more than
  $\sim1\%$ of the recently observed astrophysical neutrino flux consists of
  prompt emission from GRBs that are potentially observable by existing
  satellites.
\end{abstract}

\begin{document}

\maketitle

\section{Introduction}

While cosmic rays have been observed with energies up to
$\unit[10^{20}]{eV}$, their sources remain unknown.  Gamma-ray bursts (GRBs)
have been proposed \citep{Vietri:1995hs} as promising candidate sources of
ultra-high energy cosmic rays (UHECRs) because of their extremely large
energy release over timescales of only $\sim\unit[10^{-3}-10^3]{s}$.  In the
popular fireball model
\citep[e.g.][]{ShemiPiran1990,Piran:2004ba,Meszaros:2006rc}, gamma-rays are
produced by the dissipation of kinetic energy in an ultra-relativistic
fireball flowing outward from a cataclysmic stellar collapse or merger.  If
GRBs accelerate protons with comparable efficiency to electrons, then they
could account for most or all of the UHECR flux~\citep{waxman:1995vg}.  In
this case, protons and gamma-rays in the fireball interact through channels
such as the $\Delta$-resonance process $p+\gamma\to\Delta^+\to n+\pi^+$.
The charged pions decay leptonically via $\pi^+\to\mu^++\nu_\mu$ followed by
$\mu^+\to e^++\nu_e+\bar\nu_\mu$.  \citet{waxman:97} noted that this
neutrino flux could be measured on Earth by a sufficiently large detector.
Neutrinos correlated with GRBs would be a ``smoking-gun'' signal for UHECR
acceleration in GRBs.  To date, however, neither IceCube
\citep{Abbasi:2011qc,abbasi:12:naturegrbs} nor
ANTARES~\citep{ANTARESGRB2013} have observed such a signal.

IceCube is a $\unit{km^3}$ scale neutrino detector deployed deep in the
south polar ice cap.  The completed detector consists of 5160 digital
optical modules (DOMs), with 60 DOMs mounted on each of 86 strings.
Construction was performed during Austral summers, with the final strings
deployed in 2010 December.  Photomultiplier tubes (PMTs) in the DOMs detect
Cherenkov light emitted by energetic charged particles produced in
neutrino--nucleon interactions in the ice.  When a DOM collects sufficient
charge, digitized PMT waveforms are transmitted to the data acquisition
system (DAQ) at the surface of the ice.  When eight DOMs initiate such
launches within $\unit[5]{\mu s}$, a trigger is formed which results in
initial processing, filtering, and further transmission of data via
satellite to servers in the north.  In previous publications, the PMTs
\citep{2010:IceCube:PMT}, data acquisition methods \citep{Abbasi:2008aa},
and overall detector operations \citep{2006:IceCube:String21} have been
discussed in detail.  Datasets were collected during construction using the
partially completed detector configurations, each of which was active for
approximately one year.  The results presented here are derived from the
first year of data from the completed 86 string detector in addition to data
from the 40, 59, and 79 string configurations.

While IceCube is sensitive to neutral and charged-current interactions of
all neutrino flavors coming from any direction, in this analysis, we
restrict our focus to up-going charged-current $\nu_\mu$ interactions at
energies above \unit[1]{TeV}.  Product muons from such a signal can travel
several kilometers through the ice, providing high detection efficiency and
good angular resolution that both improve with increasing neutrino energy.
By selecting up-going muons with declination greater than $-5^\circ$, we use
the Earth (and, near the horizon, the ice cap itself) as a shield to
attenuate the large flux of muons produced by cosmic-ray interactions in the
atmosphere.  The search will be extended to all interaction channels and the
entire sky in separate papers.

\section{Data}

The originating direction of muons passing through IceCube is reconstructed
using a maximum likelihood method \citep{ahrens:04} to fit the spatial and
temporal Cherenkov light pattern observed by the DOMs.  IceCube is sensitive
to muons with sufficiently high energy that the interaction frame is highly
boosted with respect to the detector frame so that the muon trajectory is
nearly collinear with the neutrino.  Neutrino angular resolution is affected
by both the deviation angle of the product muon, which decreases with
increasing neutrino energy, and the accuracy of the reconstruction of the
muon track, which is limited by light timing uncertainties due to photon
scattering in the ice.  Including both of these effects, the median neutrino
angular error for simulated neutrinos surviving the quality cuts used in
this analysis is $1^\circ$ at $\sim$\,TeV energies; at $\sim$\,PeV energies,
this value improves to $0.5^\circ$ and the muon deviation angle is
negligible.  For each neutrino individually, the angular uncertainty
($\sigma_\nu$) is estimated using the width of the optimum in the fit
likelihood space~\citep{Neunhoffer:2004ha}.

Muon energy is reconstructed by measuring the charge collected by the DOMs
as the muon traverses the detector.  Very good neutrino energy resolution is
possible for analyses requiring the interaction vertex to be contained
within the instrumented volume \citep{Aartsen:2013vja}.  In this search,
most of the sensitivity comes from neutrinos interacting outside of the
instrumented volume.  Since the location of the interaction vertex is
generally not known, muons can lose significant energy before reaching the
instrumented volume.  Therefore, the reconstructed muon energy must be
interpreted as an approximate lower bound on the neutrino energy.

Down-going cosmic-ray-induced muons trigger the completed detector at a rate
of over \unit[2]{kHz}.  A large fraction of these events are correctly
reconstructed as down-going and are easily excluded from this analysis.  The
dominant remaining backgrounds are muons passing near the boundary of the
instrumented volume and emitting light upwards and multiple independent
muons traversing the detector at the same time.  These backgrounds, which
often yield incorrect up-going reconstructions, are rejected using
parameters described in previous work~\citep{2011:Abbasi:IC40PS} including
(1) fit quality parameters from a progression of reconstructions that apply
increasingly detailed ice and DOM response modeling; (2) comparison of the
fit quality for unbiased and down-going-biased reconstructions; (3)
reconstruction results for time- and geometry-based split subsets of the
event data; and (4) topology variables related to the distribution of DOM
pulses about the reconstructed muon path.  Event selection criteria were
optimized separately for each detector configuration.  For the 40 and 59
string configurations, previously published event selection criteria were
re-used.  For the 40 string configuration, a simple set of cuts selected
events which performed well in several quality
criteria~\citep{Abbasi:2011qc}, while for subsequent configurations, Boosted
Decision Tree forests~\citep{Freund97adecision-theoretic} were used to
synthesize a single quality parameter from all available event information.
The final sample has a data rate of $\sim$$\unit[3.8]{mHz}$ in the completed
detector and consists primarily of atmospheric muon neutrinos from the
northern hemisphere with $\sim15\%$ contamination from misreconstructed
cosmic-ray-induced muons.  Atmospheric neutrinos constitute an irreducible
background which can only be separated statistically from astrophysical
neutrinos based on reconstructed energy and temporal and directional
correlation with a GRB.

Between 2008 April 5 and 2012 May 15, 592 GRBs were observed at
declinations greater than $-5^\circ$ and reported via the GRB Coordinates
Network\footnote{http://gcn.gsfc.nasa.gov} and the Fermi GBM
catalogs~\citep{Gruber:2014iza,vonKienlin:2014nza}.  Bursts during
commissioning and calibration phases are excluded.  This analysis includes
506 bursts which occurred during stable IceCube data collection.  The search
window is determined by the time of gamma emission and the location in the
sky for each burst.  When multiple satellites observed a given burst, the
gamma emission time ($T_{100}$) is defined by the most inclusive start and
end times ($T_1$ and $T_2$) reported by any satellite.  The angular window
is determined by the direction and angular uncertainty ($\sigma_\text{GRB}$)
given by the satellite reporting the smallest angular uncertainty.  Fermi
GBM, which observes the most bursts, typically has a total statistical plus
systematic uncertainty of a few degrees or more, but for bursts observed by
other satellites, the uncertainty is generally
$\ll1^\circ$~\citep{integral,swift,superagile,HurleyEtAl2010}.  When an
asymmetric error ellipse is reported, the larger axis is used.  The small
GRB time and space windows, along with the low atmospheric neutrino rate,
make this a nearly background-free search, with a sensitivity that improves
nearly linearly with the number of bursts observed.  For modeling neutrino
fluence predictions, gamma-ray fluence parameters are taken from satellite
measurements, and unmeasured model inputs are assumed as in our previous
work~\citep{abbasi:10}.  We catalog burst information in a publically
accessible online database\footnote{http://icecube.wisc.edu/science/tools}.

\section{Analysis}

We use an unbinned maximum likelihood analysis based on~\citet{Braun:2008bg}
to test for a correlation between GRBs and neutrino events.  The likelihood
$\mathcal{S}$ that a given event is a signal event and $\mathcal{B}$ that it
is a background event are the products of separately normalized time,
direction, and energy probability distribution functions (PDFs):
\begin{align}
  \mathcal{S}/\mathcal{B}
  = (S/B)_\text{time}
  {} \ (S/B)_\text{dir}
  {} \ (S/B)_\text{energy}.
\end{align}

For a given burst, the signal time PDF is constant during gamma emission.
Before and after gamma emission, the signal time PDF falls smoothly to zero
with Gaussian tails that have a width parameter given by
\begin{equation}
  \sigma_\text{time}
  = \left\{
  \begin{array}{l r@{\,T_{100}\,}l}
    \unit[2]{s} & & < \unit[2]{s}, \\
    T_{100} & \unit[2]{s} \leq & < \unit[30]{s}, \\
    \unit[30]{s} & \unit[30]{s} \leq &.
  \end{array}
  \right.
\end{equation}
The burst time window is truncated at $4\sigma_\text{time}$ before and after
the gamma emission, and the background time PDF is constant throughout this
time window.  The signal direction PDF is a two-dimensional circular
Gaussian:
\begin{align}
  S_\text{dir}(\nu,GRB)
  = \frac{1}{2\pi\sigma_\text{dir}^2}
  {} \,\exp\Gp{-\frac{\Delta\Psi^2}{2\sigma_\text{dir}^2}},
\end{align}
where $\sigma_\text{dir}^2=\sigma_\text{GRB}^2 + \sigma_\nu^2$ and
$\Delta\Psi$ is the angular separation between the burst and the
reconstructed muon direction.  The background direction PDF is constructed
from off-time data, accounting for the declination-dependent atmospheric
neutrino event rate.  The energy PDFs are computed from the reconstructed
muon energy.  While this reconstruction only provides a lower bound on the
neutrino energy, it is nevertheless useful for probabilistically
distinguishing a possible astrophysical flux from the atmospheric
background, which has a softer spectrum.  The background energy PDF is taken
from off-time data in the energy range where we have good statistics; at
higher energies, this PDF is extended using simulated atmospheric neutrinos.
The signal energy PDF is computed using simulated signal events with an
$E^{-2}$ spectrum, which provides good sensitivity to a wide range of GRB
model spectra.

In this search, the observed number of events $N$ in the on-time window is
not known \emph{a priori}.  For supposed signal and background event rates
$n_s$ and $n_b$, respectively, the probability of observing $N$ events is
given by the Poisson distribution:
\begin{align}
  P(n_s,n_b)
  &= \frac{(n_s + n_b)^N}{N!} \exp[-(n_s+n_b)].
\end{align}
Without knowledge of the signal and background PDFs, the probabilities of an
observed event representing signal or background are $n_s/(n_s+n_b)$ and
$n_b/(n_s+n_b)$, respectively.  These probabilities are combined with the
per-event signal and background likelihoods $\mathcal{S}_i$ and
$\mathcal{B}_i$ to obtain a single likelihood for each event $i$:
\begin{align}
  \mathcal{L}_i(n_s,n_b)
  &= \frac{n_s \mathcal{S}_i + n_b \mathcal{B}_i}{n_s+n_b}.
\end{align}
The product of the Poisson probability and the per-event likelihoods give an
ensemble likelihood.  We replace the background rate hypothesis $n_b$ with
the measured rate $\Braket{n_b}$, which is well-measured in off-time data.
Because the background rate varies with detector configuration due to the
increasing size of the instrumented volume after each construction season,
an ensemble likelihood is calculated for each configuration $c$.  The
overall likelihood is a function of the per-configuration signal rates
$\GB{(n_s)_c}$ and is given by the product of the per-configuration
likelihoods:
\begin{align}
  \mathcal{L}(\GB{(n_s)_c})
  &= \prod_c P((n_s)_c) \prod_{i=1}^{N_c} \mathcal{L}_i((n_s)_c).
\end{align}
Our test statistic is the log-likelihood-ratio $T=\ln[\mathcal{L}(\GB{(\hat
n_s)_c})/\mathcal{L}(\GB{0})]$, where the values $\GB{(\hat n_s)_c}$
maximize the likelihood and $\mathcal{L}(\GB{0})$ is the likelihood for
background-only.  The test statistic can be written as
\begin{align}
  T
  = \sum_c \GB{
  {} -(\hat n_s)_c + \sum_{i=1}^{N_c} \ln \Gb{
  {} \frac{(\hat n_s)_c\,\mathcal{S}_i}{\Braket{n_b}_c\,\mathcal{B}_i} + 1 }}.
\end{align}

We use a frequentist method to derive statistical significance and fluence
upper limits from actual observations.  The significance of an observed test
statistic $T_\text{obs}$ is the probability $p$ of finding $T\geq
T_\text{obs}$ given background alone.  To find this probability,
pseudo-experiments are performed in which background-like data samples are
generated by drawing from the reconstructed energy, direction and angular
error distributions observed in off-time data.  The resulting $T$
distribution sets the significance of any single observation.  We calculate
fluence upper limits using a Feldman--Cousins
approach~\citep{feldmancousins:98}.  Simulated events weighted to a given
spectrum and normalization are added to pseudo-experiments; the exclusion
confidence level (CL) is the fraction of pseudo-experiments which yield
$T\geq T_\text{obs}$.

When expressing constraints in terms of a quasi-diffuse flux, we assume that
the 506 northern hemisphere bursts included in our four-year analysis are
representative of $n_\text{GRB}$ bursts per year that are potentially
observable by existing satellites.  Potentially observable bursts can go
unseen because they are hidden by the Sun or Moon; they occur outside the
field of view of any satellite or during satellite downtime; or, in this
analysis, because they are in the southern sky.  The extrapolation from
actually observed bursts to potentially observable bursts is uncertain due
to the differing fields of view and sensitivities of existing satellites,
but here we assume $n_\text{GRB}=667$ | the same approach used in our
previous publications~\citep{2011PhRvL.106n1101A,abbasi:12:naturegrbs}.  Our
results can be reinterpreted for a different supposed burst rate
$n_\text{GRB}^\prime$ by multiplying our reported flux values by
$n_\text{GRB}^\prime/667$.  A potentially large population of nearby,
low-luminosity GRBs~\citep{Liang:2006ci} may contribute to an observable
diffuse neutrino flux~\citep{MuraseEtAl2006}, but because they rarely
trigger gamma-ray detectors, these bursts are not directly constrained by
our analysis.

Our results are subject to systematic uncertainties in our neutrino signal
simulation.  Detector response and ice property uncertainties are accounted
for by repeating the simulation with varied values for these inputs.
Uncertainties due to muon propagation, Earth model parameters, and neutrino
interaction cross sections have been studied in detail in previous work;
these effects give a maximum uncertainty of $\sim8\%$~\citep{amandaPS}.  The
cumulative amplitude of these effects, which are included in all results
presented in this paper, is spectrum-dependent, but generally the fluence
corresponding to a given exclusion CL is increased by $\sim10\%$. 

\section{Results}

\begin{table}[t]
  \centering
  \begin{tabular}{l r r}
    \toprule
    & GRB100718A & IceCube $\nu$ \\
    \midrule
    Time
    & $T_{100}$=\unit[39]{s} & $T_1 + \unit[15]{s}$ \\
    Angular separation
    & & $16^\circ$ \\
    Angular uncertainty
    & $10.2^\circ$
    & $1.3^\circ$ \\
    GRB fluence
    & $\unit[\sci{2.5}{-6}]{erg\,cm^{-2}}$ \\
    $\nu$ energy
    & & $\gtrsim \unit[10]{TeV}$ \\
    \bottomrule
  \end{tabular}
  \caption{GRB and neutrino properties for the single coincidence observed
  in four years of data.  The quoted GRB angular uncertainty is the Fermi
  GBM statistical error for this burst.  In our analysis, the statistical
  error for GBM bursts is added in quadrature with a two-component estimated
  systematic error: $2.6^\circ$ with 72\% weight plus $10.4^\circ$ with 28\%
  weight~\citep{paciesas:12}.  No GCN circular was produced for this burst;
  however, its observation was reported in the second Fermi GBM catalog
  \citep{vonKienlin:2014nza}.  The reconstructed energy of the product muon
  is $\unit[10]{TeV}$.  As discussed above, the neutrino energy may be
  larger.}
  \label{tab:ic79track}
\end{table}

\begin{figure}[t]
  \begin{center}
    \includegraphics[width=.5\columnwidth]{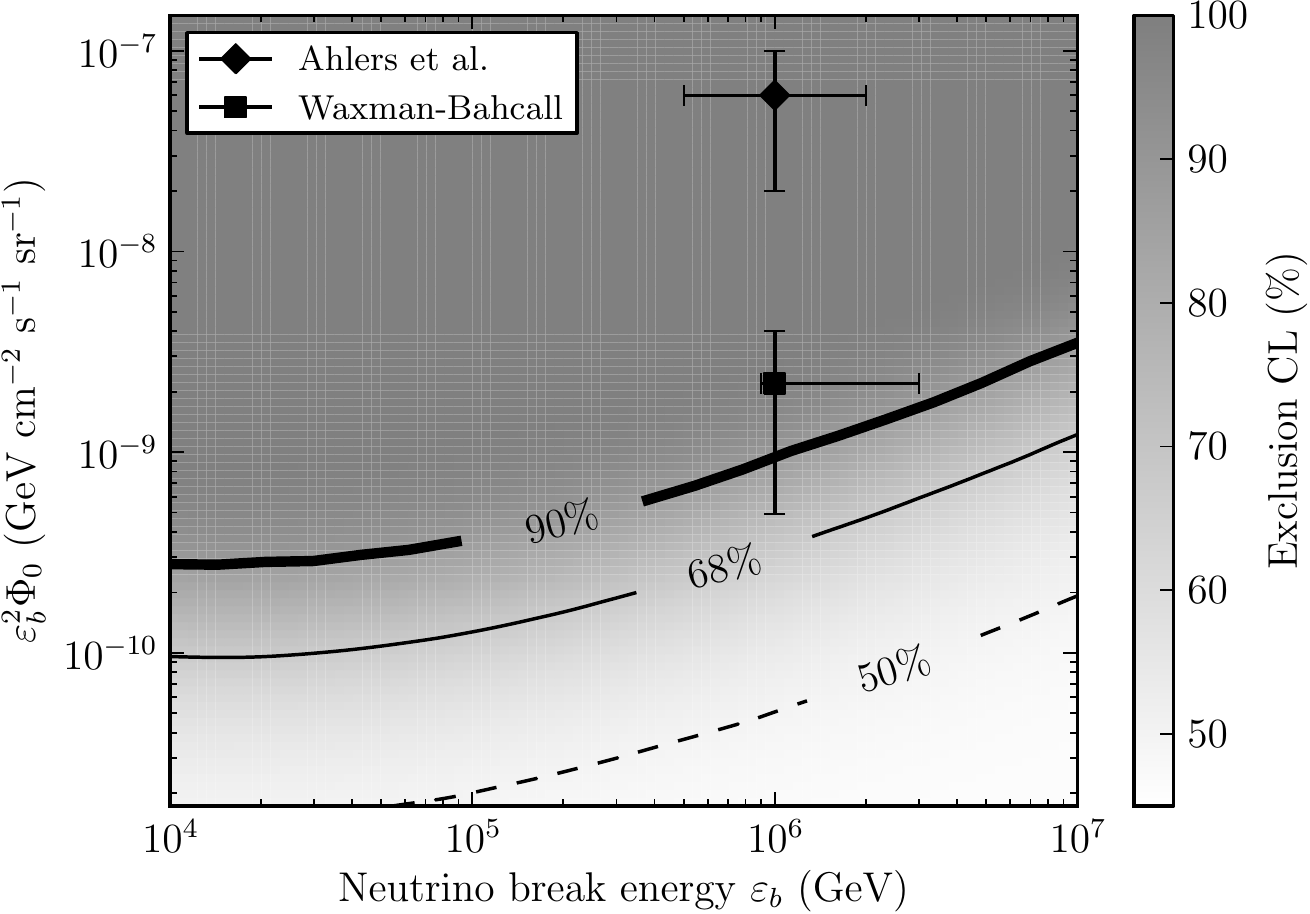}
  \end{center}
  \caption{Constraint on generic doubly broken power law neutrino flux
  models as a function of first break energy $\varepsilon_b$ and
  normalization $\Phi_0$.  The model by \citet{grbsonprobation} assumes that
  only neutrons escape from the GRB fireball to contribute to the UHECR
  flux.  The Waxman--Bahcall model \citeyearpar{waxman:97}, which allows all
  protons to escape the fireball, has been updated to account for more
  recent measurements of the UHECR flux~\citep{1475-7516-2009-03-020} and
  typical gamma break energy~\citep{goldstein:12}.}
  \label{fig:break:exclusion:CLs}
\end{figure}

In four years of data, we find a single neutrino candidate event correlated
with a GRB, yielding a significance of $p=0.46$.  The burst and neutrino
properties are listed in Table~\ref{tab:ic79track}.  Because this
observation is not significant, we are able to improve upon our previously
published upper limits \citep{abbasi:12:naturegrbs}.  First, we consider a
simple class of models for which each burst produces the same flux with a
doubly broken power law spectrum in the Earth's frame, such that the total
quasi-diffuse flux takes the form:
\begin{equation}
  \Phi_\nu (E)
  = \Phi_0 \cdot
  {} \left\{
  \begin{array}{l r@{\,E\,}l}
    E^{-1} \varepsilon_b^{-1} & & < \varepsilon_b, \\
    E^{-2} & \varepsilon_b \leq & < 10\varepsilon_b, \\
    E^{-4} (10\varepsilon_b)^{2} & 10\varepsilon_b \leq &.
  \end{array}
  \right.
\end{equation}
We show exclusion contours for such models in
Figure~\ref{fig:break:exclusion:CLs}.  Our treatment here is similar to that
in \citet{abbasi:12:naturegrbs}, but with the following modifications: (1)
the inclusion of the second spectral break at $10\varepsilon_b$, and (2) the
use of an updated Waxman-Bahcall prediction which accounts for more recent
measurements of the UHECR flux \citep{1475-7516-2009-03-020} and typical
gamma break energy~\citep{goldstein:12} in accordance with the original
prescription from~\citet{waxman:97}.  The model by \citet{grbsonprobation}
assumes that only neutrons escape from the GRB fireball to contribute to the
UHECR flux; this scenario is strongly excluded by our limit.  The
Waxman-Bahcall model allows protons to escape the fireball as UHECRs
directly without producing neutrinos, so it is not yet strongly excluded by
our observations.

In models that predict per-burst neutrino spectra based on the details of
the measured gamma-ray spectra, the fluence normalization scales linearly
with the baryonic loading $f_p=1/f_e$, where $f_e$ is the ratio of the
kinetic energy in electrons to the total energy in protons within the
fireball.  In response to our previously published model-dependent
limits~\citep{abbasi:12:naturegrbs}, \citet{Baerwald:2014zga} and others
have observed that the relevant parameter space for $f_p$ in the context of
UHECR production depends on the energy range over which the baryonic loading
is defined.  We adopt the convention that $f_p$ is defined over all proton
energies | not just energies relevant to cosmic-ray production.  Additional
modeling corrections have also been studied.  More detailed treatment of the
$p+\gamma\to\Delta^+$ process leads to a fluence reduction while the use of
numerical simulation to include other standard model $p\gamma$ interaction
channels gives a fluence enhancement~\citep{hummer:12}.

Using a wrapper for SOPHIA~\citep{2000CoPhC.124..290M} to calculate
per-burst spectra, we evaluate exclusion contours in three scenarios.  One
is the standard fireball picture~\citep{hummer:12}.  Another is a
photospheric model which moves the neutrino production to the photosphere,
where the fireball transitions from optically thick to optically thin for
$\gamma\gamma$ interactions~\citep{ReesMeszaros2005,Murase:2008sp,zhang:13}.
Finally, we consider a Poynting-dominated flux model | Internal
Collision-induced MAgnetic Reconnection and Turbulence, or
ICMART~\citep{ZhangYan2011} | in which internal shocks and particle
acceleration take place at a much higher radius, typically
$\unit[10^{16}]{cm}$ \citep{zhang:13}.

\begin{figure}[t]
  \begin{center}
    \includegraphics[width=.32\textwidth]{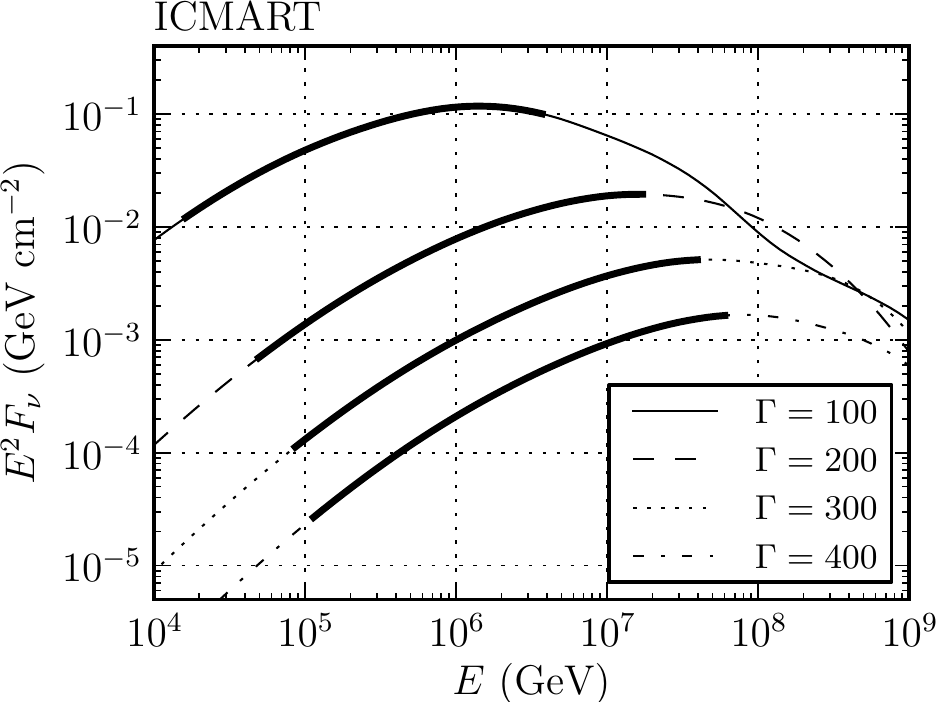}
    \hfill
    \includegraphics[width=.32\textwidth]{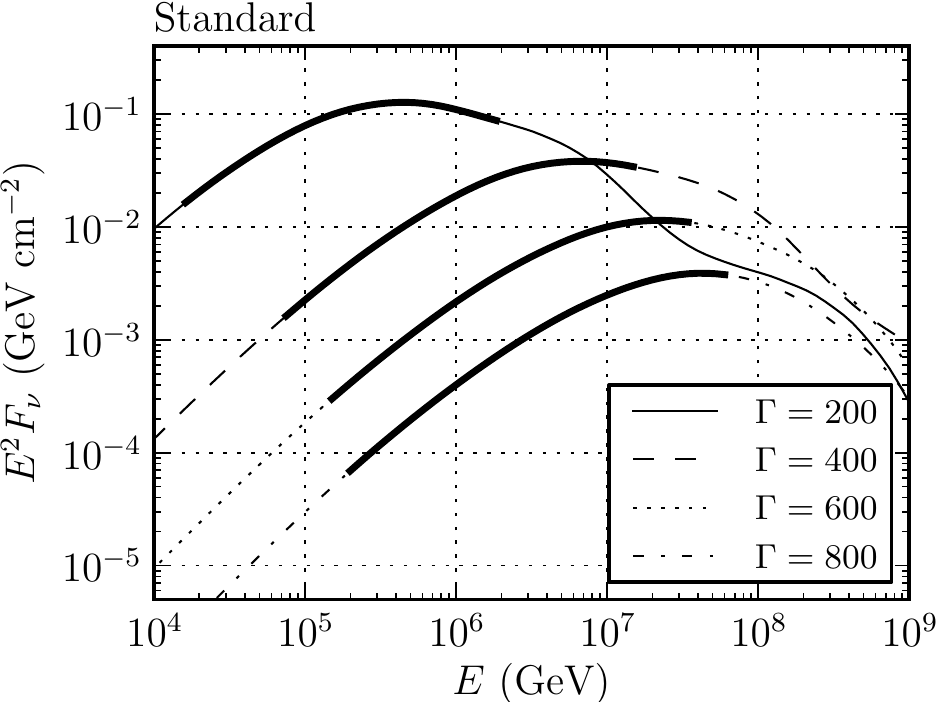}
    \hfill
    \includegraphics[width=.32\textwidth]{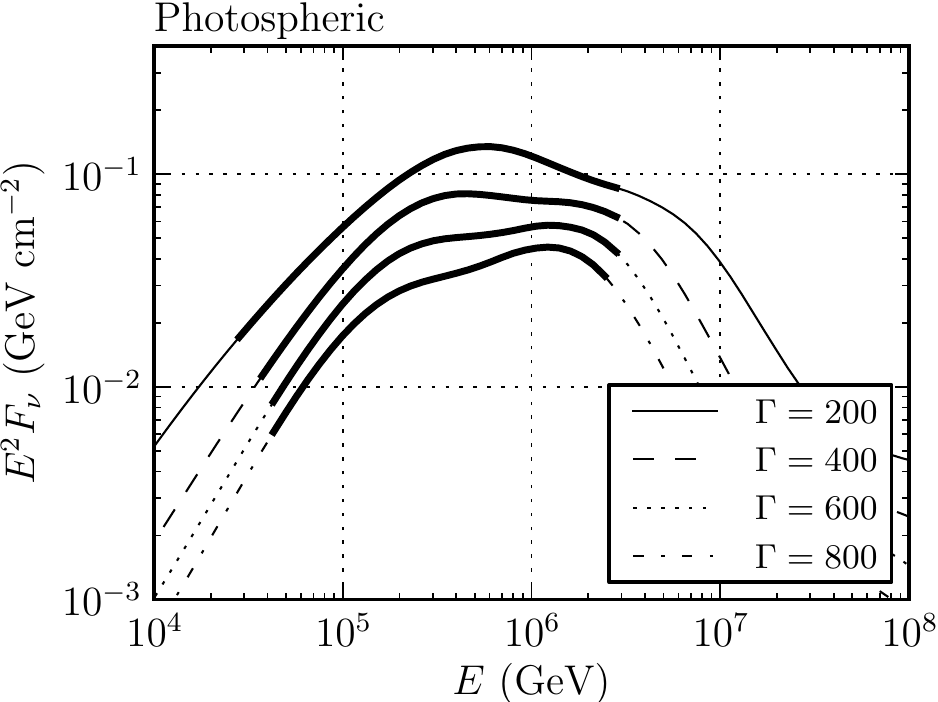}
  \end{center}
  \caption{Total predicted neutrino fluence for various values of the bulk
  Lorentz factor $\Gamma$ under different model assumptions.  Bold lines
  reflect the energy region in which 90\% of events are expected based on
  simulation.  Normalization scales linearly with the assumed baryonic
  loading $f_p$, which is set here to 10.  Models are arranged from left to
  right in order of increasing predicted fluence for given values of $f_p$
  and $\Gamma$.}
  \label{fig:predictions}
\end{figure}

\begin{figure}[t]
  \begin{center}
    \includegraphics[width=.32\textwidth]{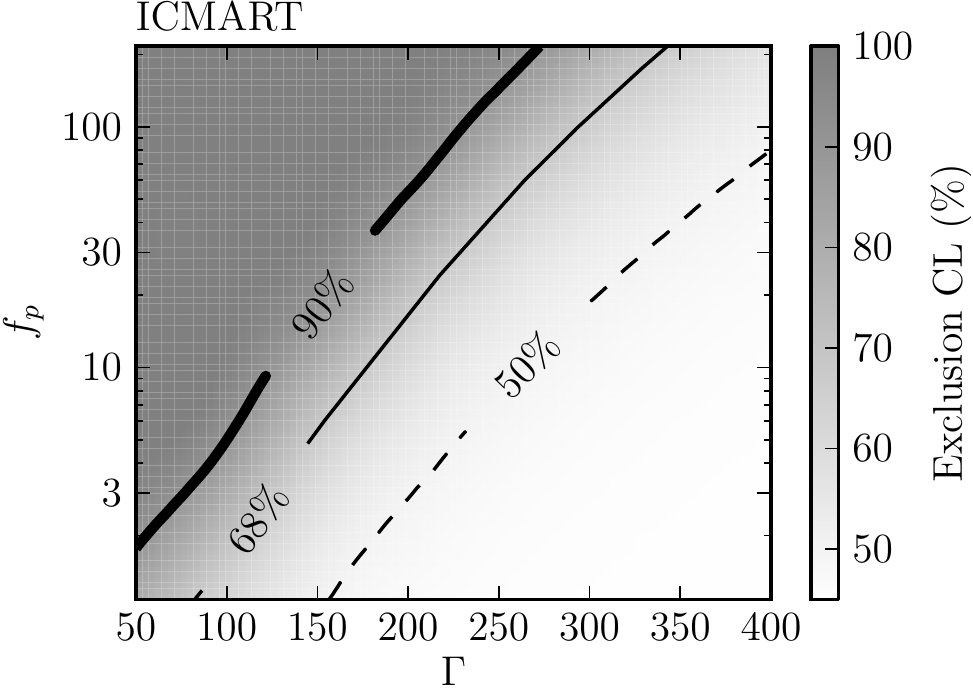}
    \hfill
    \includegraphics[width=.32\textwidth]{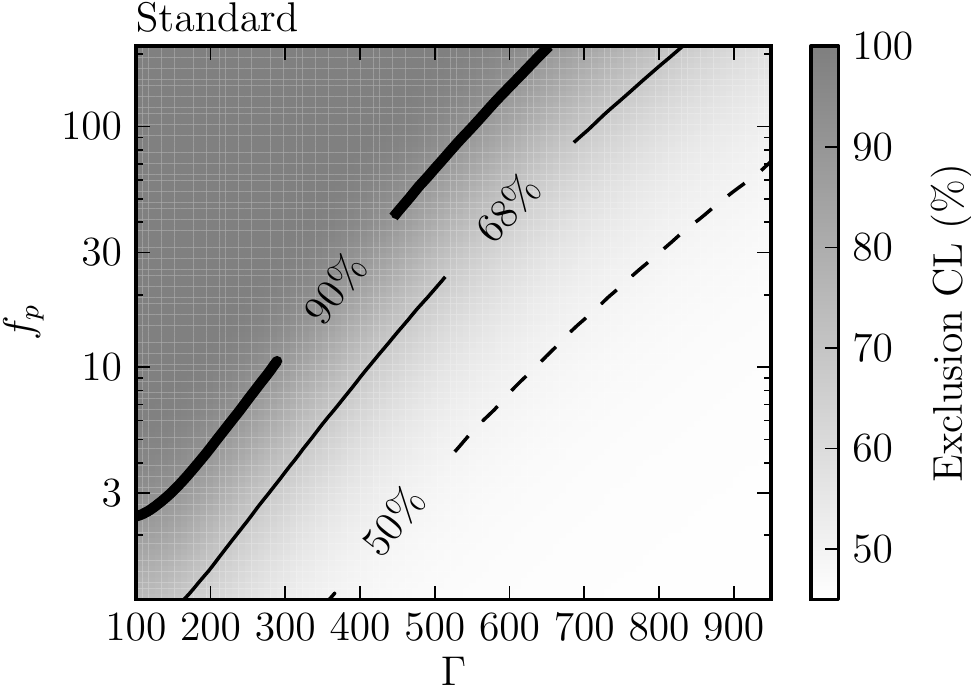}
    \hfill
    \includegraphics[width=.32\textwidth]{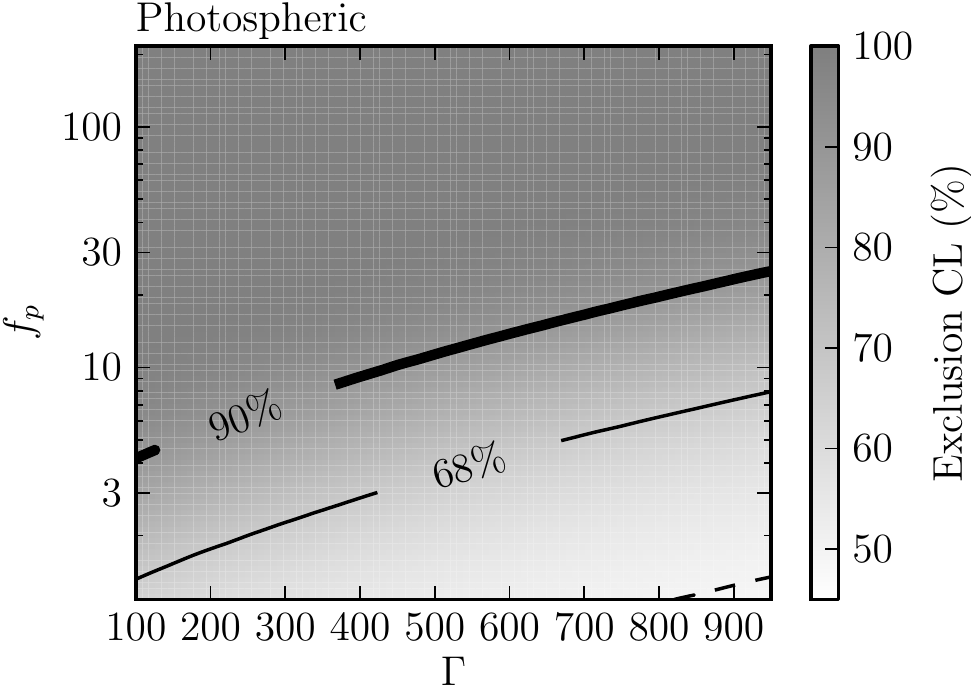}
  \end{center}
  \caption{Allowed region for the baryonic loading $f_p$ and bulk Lorentz
  factor $\Gamma$ under different model assumptions.}
  \label{fig:model:limits}
\end{figure}

For each model, we scan the parameter space for the bulk Lorentz factor of
the fireball $\Gamma$ and the baryonic loading $f_p=1/f_e$.  In each case,
we consider $1 < f_p < 200$.  For the standard and photospheric models, we
test $100<\Gamma<950$ while for ICMART, which varies more strongly with
$\Gamma$, we test $50<\Gamma<400$.  The predicted spectra, summed over all
analyzed bursts, are shown in Figure~\ref{fig:predictions}; the resulting
exclusion contours are shown in Figure~\ref{fig:model:limits}.  Our results
rule out some of the parameter space for $f_p$ and $\Gamma$ in regions that
allow GRBs to be dominant UHECR sources.  For very large values of $\Gamma$,
IceCube would require a very long exposure to constrain the models.
However, this region can be probed in other ways, such as by improved energy
calibration of cosmic-ray measurements~\citep{Baerwald:2014zga}.   We note
that the constraints calculated here do not account for a possible
enhancement to the high energy neutrino flux due to acceleration of
secondary particles~\citep{Winter:2014tta} or a distribution of differing
$\Gamma$~\citep{2012ApJ_He}; nor do we attempt to account for
a possible reduction of the neutrino flux if the brightest GRBs (in
gamma-rays) have a smaller baryonic loading~\citep{Asano:2014nba}.

IceCube has recently established~\citep{Aartsen:2014gkd,Aartsen:2014muf} the
existence of an astrophysical neutrino flux whose sources, like those of the
UHECRs, are not yet known.  This flux is established by neutrino events
above expected backgrounds in the \unit[10]{TeV} to few PeV range.  The
observed signal is consistent with an isotropic flux and can be
parameterized as $\Phi_\nu(E) = \Phi_0 (E/E_0)^{-\gamma}$.  If $E_0$ is
taken to be \unit[100]{TeV}, then the best fit gives a per-flavor
$\nu+\bar\nu$ normalization $E_0^2\Phi_0=\sci{2.06^{+0.4}_{-0.3}}{-8}\,
\unit{GeV\,cm^{-2}\,s^{-1}\,sr^{-1}}$ and spectral index
$\gamma=2.46\pm0.12$~\citep{Aartsen:2014muf}.  To constrain the contribution
to this flux from GRBs, we follow the prescription applied above for
doubly broken power law spectra, except this time the simulation is weighted
to unbroken spectra with $2<\gamma<2.6$.  Only simulated events above
\unit[10]{TeV} are considered; at very high energies, where the flux is
already much smaller, no explicit cutoff is made.  We find that the allowed
GRB per-flavor $\nu+\bar\nu$ normalization, at 90\% CL, is
$E_0^2\Phi_0\sim\sci{2}{-10}\, \unit{GeV\,cm^{-2}\,s^{-1}\,sr^{-1}}$.  This
constraint weakens only slightly with increasing $\gamma$.  Thus potentially
observable GRBs, as defined in this paper, contribute no more than $\sim1\%$
of the observed diffuse flux.

In this work, we have only considered a handful of possible neutrino
spectra.  In recognition of the large space of possible models to test, we
now provide an online tool for calculating limits on alternative spectra.
The subset of analyzed bursts to include as well as the per-burst spectra
must be provided by the user.  These choices are applied to our full
analysis chain, and the results are sent back to the user via e-mail.
Calculating limits in this way accounts for the details of our unbinned
likelihood analysis, most importantly including the energy PDF; it also
accounts for the one low-significance event which has been observed so far.
See \texttt{http://icecube.wisc.edu/science/tools} for more details.

\section{Conclusion}

Using four years of IceCube data, we set the most stringent limits yet on
GRB neutrino production, with a sensitivity improvement of $\sim2\times$
relative to our previous results.  We constrain parts of the parameter space
relevant to the production of UHECRs in the latest models.  In addition to
the work presented here, complementary analyses are underway.  We are
improving our acceptance with a search in the cascade channel, which is
sensitive to the whole sky and to all neutrino interactions other than muon
charged-current, as well a search for GRB-correlated high energy starting
events, which has an extremely low background rate and therefore is
sensitive to very early precursor or late afterglow neutrinos.  Results from
these searches will soon be published separately.  In the absence of an
emerging signal in the coming years, IceCube limits will increasingly
constrain GRBs as dominant sources of UHECRs.

\acknowledgments

We acknowledge the support from the following agencies:
U.S. National Science Foundation-Office of Polar Programs,
U.S. National Science Foundation-Physics Division,
University of Wisconsin Alumni Research Foundation,
the Grid Laboratory Of Wisconsin (GLOW) grid infrastructure at the
University of Wisconsin - Madison, the Open Science Grid (OSG) grid
infrastructure;
U.S. Department of Energy, and National Energy Research Scientific Computing
Center,
the Louisiana Optical Network Initiative (LONI) grid computing resources;
Natural Sciences and Engineering Research Council of Canada,
WestGrid and Compute/Calcul Canada;
Swedish Research Council,
Swedish Polar Research Secretariat,
Swedish National Infrastructure for Computing (SNIC),
and Knut and Alice Wallenberg Foundation, Sweden;
German Ministry for Education and Research (BMBF),
Deutsche Forschungsgemeinschaft (DFG),
Helmholtz Alliance for Astroparticle Physics (HAP),
Research Department of Plasmas with Complex Interactions (Bochum), Germany;
Fund for Scientific Research (FNRS-FWO),
FWO Odysseus programme,
Flanders Institute to encourage scientific and technological research in
industry (IWT),
Belgian Federal Science Policy Office (Belspo);
University of Oxford, United Kingdom;
Marsden Fund, New Zealand;
Australian Research Council;
Japan Society for Promotion of Science (JSPS);
the Swiss National Science Foundation (SNSF), Switzerland;
National Research Foundation of Korea (NRF);
Danish National Research Foundation, Denmark (DNRF)


\begin{thebibliography}{}
\expandafter\ifx\csname natexlab\endcsname\relax\def\natexlab#1{#1}\fi

\bibitem[{Aartsen {et~al.}(2015)Aartsen, Ackermann, Adams, Aguilar, Ahlers,
  {et~al.}}]{Aartsen:2014muf}
Aartsen, M., Ackermann, M., Adams, J., {et~al.} 2015, PhRvD, D91, 022001

\bibitem[{Aartsen {et~al.}(2014{\natexlab{a}})Aartsen, Ackermann, Adams,
  {et~al.}}]{Aartsen:2013vja}
Aartsen, M., Ackermann, M., Adams, J., {et~al.} 2014{\natexlab{a}}, JInst, 9,
  P03009

\bibitem[{Aartsen {et~al.}(2014{\natexlab{b}})Aartsen, Ackermann, Adams,
  {et~al.}}]{Aartsen:2014gkd}
Aartsen, M., Ackermann, M., Adams, J., {et~al.} 2014{\natexlab{b}}, PhRvL, 113,
  101101

\bibitem[{Abbasi {et~al.}(2010)Abbasi, Abdou, Abu-Zayyad, {et~al.}}]{abbasi:10}
Abbasi, R., Abdou, Y., Abu-Zayyad, T., {et~al.} 2010, ApJ, 710, 346

\bibitem[{{Abbasi} {et~al.}(2011{\natexlab{a}}){Abbasi}, {Abdou}, {Abu-Zayyad},
  {et~al.}}]{2011PhRvL.106n1101A}
{Abbasi}, R., {Abdou}, Y., {Abu-Zayyad}, T., {et~al.} 2011{\natexlab{a}},
  PhRvL, 106, 141101

\bibitem[{Abbasi {et~al.}(2012)Abbasi, {Abdou}, {Abu-Zayyad},
  {et~al.}}]{abbasi:12:naturegrbs}
Abbasi, R., {Abdou}, Y., {Abu-Zayyad}, T., {et~al.} 2012, Natur, 484, 351

\bibitem[{Abbasi {et~al.}(2009)Abbasi, Ackermann, Adams,
  {et~al.}}]{Abbasi:2008aa}
Abbasi, R., Ackermann, M., Adams, J., {et~al.} 2009, NucIM, A601, 294

\bibitem[{{Abbasi} {et~al.}(2010){Abbasi}, {Abdou}, {Abu-Zayyad}, {Adams},
  {Aguilar}, {Ahlers}, {Andeen}, {Auffenberg}, {Bai}, {Baker}, \&
  et~al.}]{2010:IceCube:PMT}
{Abbasi}, R., {Abdou}, Y., {Abu-Zayyad}, T., {et~al.} 2010, NIMPA, 618, 139

\bibitem[{{Abbasi} {et~al.}(2011{\natexlab{b}}){Abbasi}, {Abdou}, {Abu-Zayyad},
  {Adams}, {Aguilar}, {Ahlers}, {Andeen}, {Auffenberg}, {Bai}, {Baker},
  {et~al.}}]{Abbasi:2011qc}
{Abbasi}, R., {Abdou}, Y., {Abu-Zayyad}, T., {et~al.} 2011{\natexlab{b}},
  PhRvL, 106, 141101

\bibitem[{{Abbasi} {et~al.}(2011{\natexlab{c}}){Abbasi}, {Abdou}, {Abu-Zayyad},
  {Adams}, {Aguilar}, {Ahlers}, {Andeen}, {Auffenberg}, {Bai}, {Baker}, \&
  et~al.}]{2011:Abbasi:IC40PS}
{Abbasi}, R., {Abdou}, Y., {Abu-Zayyad}, T., {et~al.} 2011{\natexlab{c}}, \apj,
  732, 18

\bibitem[{{Achterberg} {et~al.}(2006){Achterberg}, {Ackermann}, {Adams},
  {Ahrens}, {Andeen}, {Atlee}, {Baccus}, {Bahcall}, {Bai}, \&
  et~al.}]{2006:IceCube:String21}
{Achterberg}, A., {Ackermann}, M., {Adams}, J., {et~al.} 2006, APh, 26, 155

\bibitem[{{Achterberg} {et~al.}(2007){Achterberg}, {Ackermann}, {Adams},
  {Ahrens}, {Andeen}, {Atlee}, {Bahcall}, {Bai}, {Baret}, {Barwick}, \&
  et~al.}]{amandaPS}
{Achterberg}, A., {Ackermann}, M., {Adams}, J., {et~al.} 2007, PhRvD, 75,
  102001

\bibitem[{{Adri{\'a}n-Mart{\'{\i}}nez}
  {et~al.}(2013){Adri{\'a}n-Mart{\'{\i}}nez}, {Albert}, {Samarai}, {Andr{\'e}},
  {Anghinolfi}, {Anton}, {Anvar}, {Ardid}, {Astraatmadja}, {Aubert}, {Baret},
  {Barrios-Marti}, {Basa}, {Bertin}, {Biagi}, {Bigongiari}, {Bogazzi},
  {Bouhou}, {Bouwhuis}, {Brunner}, {Busto}, {Capone}, {Caramete},
  {C{\^a}rloganu}, {Carr}, {Cecchini}, {Charif}, {Charvis}, {Chiarusi},
  {Circella}, {Classen}, {Coniglione}, {Core}, {Costantini}, {Coyle},
  {Creusot}, {Curtil}, {De Bonis}, {Dekeyser}, {Deschamps}, {Distefano},
  {Donzaud}, {Dornic}, {Dorosti}, {Drouhin}, {Dumas}, {Eberl}, {Emanuele},
  {Enzenh{\"o}fer}, {Ernenwein}, {Escoffier}, {Fehn}, {Fermani}, {Flaminio},
  {Folger}, {Fritsch}, {Fusco}, {Galat{\`a}}, {Gay}, {Gei{\ss}els{\"o}der},
  {Geyer}, {Giacomelli}, {Giordano}, {Gleixner}, {G{\'o}mez-Gonz{\'a}lez},
  {Graf}, {Guillard}, {van Haren}, {Heijboer}, {Hello}, {Hern{\'a}ndez-Rey},
  {Herold}, {H{\"o}{\ss}l}, {James}, {de Jong}, {Kadler}, {Kalekin}, {Kappes},
  {Katz}, {Kooijman}, {Kouchner}, {Kreykenbohm}, {Kulikovskiy}, {Lahmann},
  {Lambard}, {Lambard}, {Larosa}, {Lef{\`e}vre}, {Leonora}, {Lo Presti},
  {Loehner}, {Loucatos}, {Louis}, {Mangano}, {Marcelin}, {Margiotta},
  {Mart{\'{\i}}nez-Mora}, {Martini}, {Michael}, {Montaruli}, {Morganti},
  {M{\"u}ller}, {Neff}, {Nezri}, {Palioselitis}, {P{\u a}v{\u a}la{\c s}},
  {Perrina}, {Piattelli}, {Popa}, {Pradier}, {Racca}, {Riccobene}, {Richter},
  {Rivi{\`e}re}, {Robert}, {Roensch}, {Rostovtsev}, {Samtleben}, {Sanguineti},
  {Schmid}, {Schnabel}, {Schulte}, {Sch{\"u}ssler}, {Seitz}, {Shanidze},
  {Sieger}, {Simeone}, {Spies}, {Spurio}, {Steijger}, {Stolarczyk},
  {S{\'a}nchez-Losa}, {Taiuti}, {Tamburini}, {Tayalati}, {Trovato}, {Vallage},
  {Vall{\'e}e}, {Van Elewyck}, {Vernin}, {Visser}, {Wagner}, {Wilms}, {de
  Wolf}, {Yatkin}, {Yepes}, {Zornoza}, {Z{\'u}{\~n}iga}, \&
  {Baerwald}}]{ANTARESGRB2013}
{Adri{\'a}n-Mart{\'{\i}}nez}, S., {Albert}, A., {Samarai}, I.~A., {et~al.}
  2013, A\&A, 559, A9

\bibitem[{Ahlers {et~al.}(2011)Ahlers, Gonzalez-Garcia, \&
  Halzen}]{grbsonprobation}
Ahlers, M., Gonzalez-Garcia, M., \& Halzen, F. 2011, APh, 35, 87

\bibitem[{Ahrens {et~al.}(2004)Ahrens, Bai, Bay, {et~al.}}]{ahrens:04}
Ahrens, J., Bai, X., Bay, R., {et~al.} 2004, NIMPA, 524, 169

\bibitem[{Asano \& M\'esz\'aros(2014)}]{Asano:2014nba}
Asano, K., \& M\'esz\'aros, P. 2014, ApJ, 785, 54

\bibitem[{Baerwald {et~al.}(2014)Baerwald, Bustamante, \&
  Winter}]{Baerwald:2014zga}
Baerwald, P., Bustamante, M., \& Winter, W. 2014, APh, 62, 66

\bibitem[{Braun {et~al.}(2008)Braun, Dumm, De~Palma, Finley, Karle,
  {et~al.}}]{Braun:2008bg}
Braun, J., Dumm, J., De~Palma, F., {et~al.} 2008, APh, 29, 299

\bibitem[{Feldman \& Cousins(1998)}]{feldmancousins:98}
Feldman, G., \& Cousins, R. 1998, PhRvD, 57, 3873

\bibitem[{{Feroci} {et~al.}(2007){Feroci}, {Costa}, {Soffitta}, {Del Monte},
  {di Persio}, {Donnarumma}, {Evangelista}, {Frutti}, {Lapshov}, {Lazzarotto},
  {Mastropietro}, {Morelli}, {Pacciani}, {Porrovecchio}, {Rapisarda}, {Rubini},
  {Tavani}, \& {Argan}}]{superagile}
{Feroci}, M., {Costa}, E., {Soffitta}, P., {et~al.} 2007, NIMA, 581, 728

\bibitem[{Freund \& Schapire(1997)}]{Freund97adecision-theoretic}
Freund, Y., \& Schapire, R.~E. 1997, J.~Comput.~Syst.~Sci., 55, 119

\bibitem[{{Gehrels} {et~al.}(2004){Gehrels}, {Chincarini}, {Giommi}, {Mason},
  {Nousek}, {Wells}, {White}, {Barthelmy}, {Burrows}, {Cominsky}, {Hurley},
  {Marshall}, {M{\'e}sz{\'a}ros}, {Roming}, {Angelini}, {Barbier}, {Belloni},
  {Campana}, {Caraveo}, {Chester}, {Citterio}, {Cline}, {Cropper}, {Cummings},
  {Dean}, {Feigelson}, {Fenimore}, {Frail}, {Fruchter}, {Garmire}, {Gendreau},
  {Ghisellini}, {Greiner}, {Hill}, {Hunsberger}, {Krimm}, {Kulkarni}, {Kumar},
  {Lebrun}, {Lloyd-Ronning}, {Markwardt}, {Mattson}, {Mushotzky}, {Norris},
  {Osborne}, {Paczynski}, {Palmer}, {Park}, {Parsons}, {Paul}, {Rees},
  {Reynolds}, {Rhoads}, {Sasseen}, {Schaefer}, {Short}, {Smale}, {Smith},
  {Stella}, {Tagliaferri}, {Takahashi}, {Tashiro}, {Townsley}, {Tueller},
  {Turner}, {Vietri}, {Voges}, {Ward}, {Willingale}, {Zerbi}, \&
  {Zhang}}]{swift}
{Gehrels}, N., {Chincarini}, G., {Giommi}, P., {et~al.} 2004, \apj, 611, 1005

\bibitem[{Goldstein {et~al.}(2012)Goldstein, Burgess, Preece,
  {et~al.}}]{goldstein:12}
Goldstein, A., Burgess, J., Preece, R., {et~al.} 2012, ApJS, 199, 19

\bibitem[{Gruber {et~al.}(2014)Gruber, Goldstein, Weller~von Ahlefeld, Bhat,
  Bissaldi, {et~al.}}]{Gruber:2014iza}
Gruber, D., Goldstein, A., Weller~von Ahlefeld, V., {et~al.} 2014, ApJS, 211,
  12

\bibitem[{{He} {et~al.}(2012){He}, {Liu}, {Wang}, {Nagataki}, {Murase}, \&
  {Dai}}]{2012ApJ_He}
{He}, H.-N., {Liu}, R.-Y., {Wang}, X.-Y., {et~al.} 2012, \apj, 752, 29

\bibitem[{{H{\"u}mmer} {et~al.}(2012){H{\"u}mmer}, {Baerwald}, \&
  {Winter}}]{hummer:12}
{H{\"u}mmer}, S., {Baerwald}, P., \& {Winter}, W. 2012, PhRvL, 108, 231101

\bibitem[{{Hurley} {et~al.}(2010){Hurley}, {Golenetskii}, {Aptekar}, {Mazets},
  {Pal'Shin}, {Frederiks}, {Mitrofanov}, {Golovin}, {Litvak}, {Sanin},
  {Boynton}, {Fellows}, {Harshman}, {Starr}, {Smith}, {Wigger}, {Hajdas}, {von
  Kienlin}, {Rau}, {Yamaoka}, {Ohno}, {Takahashi}, {Fukazawa}, {Tashiro},
  {Terada}, {Murakami}, {Makishima}, {Barthelmy}, {Cline}, {Cummings},
  {Gehrels}, {Krimm}, {Goldsten}, {Del Monte}, {Feroci}, {Marisaldi}, {Briggs},
  {Connaughton}, \& {Meegan}}]{HurleyEtAl2010}
{Hurley}, K., {Golenetskii}, S., {Aptekar}, R., {et~al.} 2010, in American
  Institute of Physics Conference Series, Vol. 1279, Deciphering the Ancient
  Universe with Gamma-Ray Bursts, ed. N.~Kawai \& S.~Nagataki, 330--333

\bibitem[{Katz {et~al.}(2009)Katz, Budnik, \& Waxman}]{1475-7516-2009-03-020}
Katz, B., Budnik, R., \& Waxman, E. 2009, JCAP, 2009, 020

\bibitem[{Liang {et~al.}(2007)Liang, Zhang, \& Dai}]{Liang:2006ci}
Liang, E., Zhang, B., \& Dai, Z. 2007, ApJ, 662, 1111

\bibitem[{M\'esz\'aros(2006)}]{Meszaros:2006rc}
M\'esz\'aros, P. 2006, RPPh, 69, 2259

\bibitem[{{M{\"u}cke} {et~al.}(2000){M{\"u}cke}, {Engel}, {Rachen},
  {Protheroe}, \& {Stanev}}]{2000CoPhC.124..290M}
{M{\"u}cke}, A., {Engel}, R., {Rachen}, J.~P., {Protheroe}, R.~J., \& {Stanev},
  T. 2000, CoPhC, 124, 290

\bibitem[{Murase(2008)}]{Murase:2008sp}
Murase, K. 2008, PhRvD, D78, 101302

\bibitem[{Murase {et~al.}(2006)Murase, Ioka, Nagataki, \&
  Nakamura}]{MuraseEtAl2006}
Murase, K., Ioka, K., Nagataki, S., \& Nakamura, T. 2006, ApJL, 651, L5

\bibitem[{Neunh{\"o}ffer(2006)}]{Neunhoffer:2004ha}
Neunh{\"o}ffer, T. 2006, APh, 25, 220

\bibitem[{Paciesas {et~al.}(2012)Paciesas, Meegan, von Kienlin,
  {et~al.}}]{paciesas:12}
Paciesas, W., Meegan, C., von Kienlin, A., {et~al.} 2012, ApJS, 199, 18

\bibitem[{Piran(2004)}]{Piran:2004ba}
Piran, T. 2004, RvMP, 76, 1143

\bibitem[{Rees \& M\'esz\'aros(2005)}]{ReesMeszaros2005}
Rees, M.~J., \& M\'esz\'aros, P. 2005, ApJ, 628, 847

\bibitem[{{Shemi} \& {Piran}(1990)}]{ShemiPiran1990}
{Shemi}, A., \& {Piran}, T. 1990, \apjl, 365, L55

\bibitem[{Vietri(1995)}]{Vietri:1995hs}
Vietri, M. 1995, ApJ, 453, 883

\bibitem[{von Kienlin {et~al.}(2014)von Kienlin, Meegan, Paciesas, Bhat,
  Bissaldi, {et~al.}}]{vonKienlin:2014nza}
von Kienlin, A., Meegan, C.~A., Paciesas, W.~S., {et~al.} 2014, ApJS, 211, 13

\bibitem[{Waxman(1995)}]{waxman:1995vg}
Waxman, E. 1995, PhRvL, 75, 386

\bibitem[{Waxman \& Bahcall(1997)}]{waxman:97}
Waxman, E., \& Bahcall, J. 1997, PhRvL, 78, 2292

\bibitem[{{Winkler} {et~al.}(2003){Winkler}, {Courvoisier}, {Di Cocco},
  {Gehrels}, {Gim{\'e}nez}, {Grebenev}, {Hermsen}, {Mas-Hesse}, {Lebrun},
  {Lund}, {Palumbo}, {Paul}, {Roques}, {Schnopper}, {Sch{\"o}nfelder},
  {Sunyaev}, {Teegarden}, {Ubertini}, {Vedrenne}, \& {Dean}}]{integral}
{Winkler}, C., {Courvoisier}, T., {Di Cocco}, G., {et~al.} 2003, \aap, 411, L1

\bibitem[{Winter {et~al.}(2014)Winter, Tjus, \& Klein}]{Winter:2014tta}
Winter, W., Tjus, J.~B., \& Klein, S.~R. 2014, A\&A, 569, A58

\bibitem[{Zhang \& Kumar(2013)}]{zhang:13}
Zhang, B., \& Kumar, P. 2013, PhRvL, 110, 121101

\bibitem[{Zhang \& Yan(2011)}]{ZhangYan2011}
Zhang, B., \& Yan, H. 2011, ApJ, 726, 90

\end{thebibliography}

\end{document}